\documentclass[article, twocolumn,letterpaper,groupedaddress,amsmath,amssymb,showpacs,english,superscriptaddress]{revtex4-2}

\usepackage{graphicx}
\usepackage{dcolumn}
\usepackage{bm}
\usepackage[table]{xcolor}
\usepackage{hyperref}
\usepackage{braket}
\usepackage{placeins}
\usepackage{physics}
\usepackage{amsmath} 
\usepackage{array} 
\usepackage{mathtools}
\usepackage{soul}
\usepackage[capitalize]{cleveref}
\usepackage[mathlines]{lineno}
\allowdisplaybreaks
\newcolumntype{C}[1]{>{\centering\arraybackslash}p{#1}}
\usepackage{upgreek}

\usepackage{booktabs} 
\usepackage{tabularx} 
\definecolor{lightgray}{gray}{0.9}
\definecolor{awesome}{rgb}{1.0, 0.13, 0.32}
\makeatletter

\begin{document}

\title{Distributed Quantum Computing in Silicon}
\author{Photonic Inc.}
\thanks{See \cref{sec:authors} for list of contributors.}
\date{\today}

\begin{abstract}
Commercially impactful quantum algorithms such as quantum chemistry and Shor's algorithm require a number of qubits and gates far beyond the capacity of any existing quantum processor. Distributed architectures, which scale horizontally by networking modules, provide a route to commercial utility and will eventually surpass the capability of any single quantum computing module. Such processors consume remote entanglement distributed between modules to realize distributed quantum logic. Networked quantum computers will therefore require the capability to rapidly distribute high fidelity entanglement between modules.
Here we present preliminary demonstrations of some key distributed quantum computing protocols on silicon T centres in isotopically-enriched silicon. We demonstrate the distribution of entanglement between modules and consume it to apply a teleported gate sequence, establishing a proof-of-concept for T centres as a distributed quantum computing and networking platform.
\end{abstract} 

\maketitle

\section{Introduction}

The commercial interest in quantum information is motivated in large part by the promise that, in due course, quantum technologies will be able to outperform classical computing and communication capabilities in a small number of socially and commercially important tasks. There exist known quantum computing algorithms whose computational complexity allows for super-polynomial speedups in, for example, cryptanalysis (Shor’s algorithm) and chemistry simulations (Quantum Phase Estimation, or QPE, algorithms) relative to their classical counterparts. To execute such algorithms it is expected that, at a minimum, hundreds to thousands of fault-tolerant logical qubits with logical error rates (LER) on the order of $10^{-12}$ and even beyond will be required~\cite{Vaschillo.2022}. It has been suggested that this scale of quantum resources will not be available in a single monolithic architecture for most quantum platforms under consideration \cite{GQI}. Even beyond this scale, more qubits will always be better: to either make the same algorithms run faster through space-time resource trade-offs, or to solve ever larger problems of interest. Eventually, the largest, most powerful quantum computers will be horizontally scalable through a networked, modular construction.

We propose that there will be three phases of quantum technology development on the path to commercial applications. The first, Phase 1 Quantum, already goes by the name ``Noisy Intermediate-Scale Quantum'' (NISQ). Phase 1 Quantum computing is the technological phase where single-module quantum hardware can run circuits and small algorithms, however their quantum operations are too error-prone to perform quantum error correction (QEC). We are presently witnessing the birth of Phase 2 Quantum, where single quantum modules can demonstrate QEC protocols such as surface code \cite{Acharya2022GoogleSurface} or quantum low-density parity-check (QLDPC) codes~\cite{Bluvstein_2023, Dasilva2024}.  There may be useful scientific results to emerge from these single-module computers, however many known algorithms of commercial relevance will require a number of qubits that exceeds the upper bounds for individual modules projected by most quantum hardware platforms, particularly when fault-tolerance needs are taken into account \cite{GQI}. Phase 3 Quantum is the era of quantum supercomputers, including the era of networked quantum computing, which enables horizontal scaling through fault-tolerant gates implemented between distributed logical processors. Phase 3 Quantum computers will unlock not only the commercial quantum computing applications described above, distributed quantum computing will also unlock quantum repeaters and fault-tolerant quantum networking applications like global entanglement distribution protocols and blind quantum computing.

\begin{figure}[t!]
\includegraphics[width=\columnwidth]{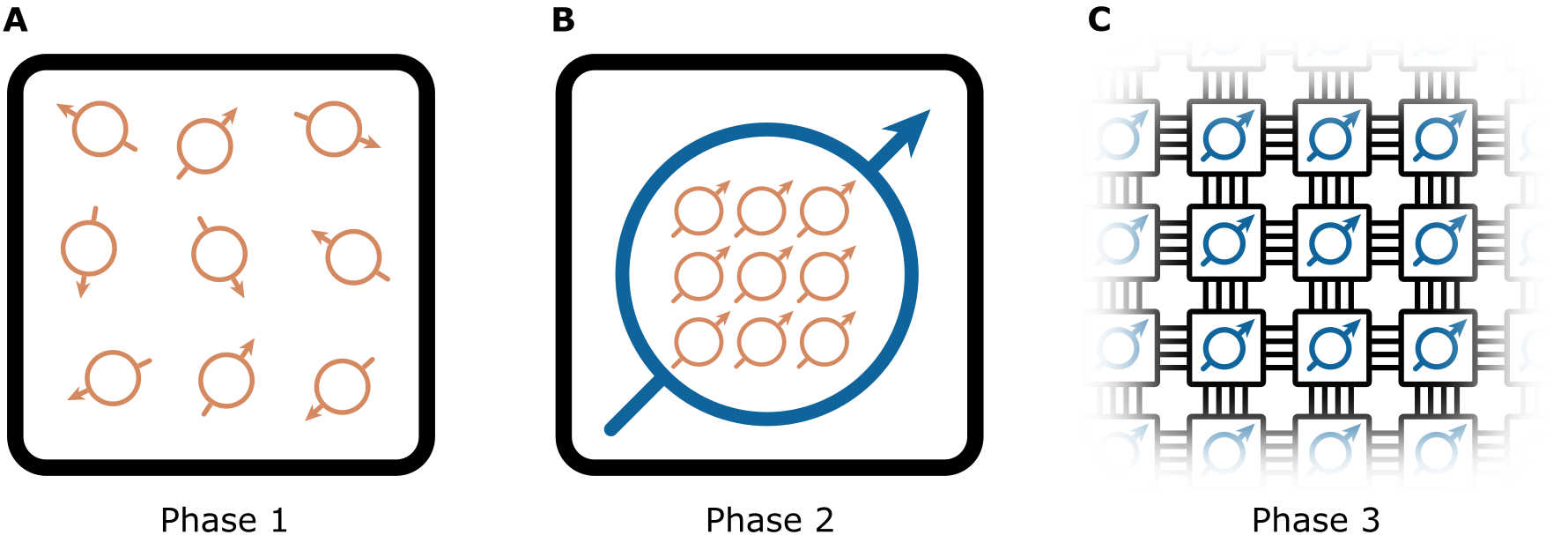}
\caption{\textbf{Phases of quantum computing}. \textbf{A.} In phase 1 Quantum computing (the NISQ phase), quantum computers are single modules containing a small number of physical qubits (orange) too noisy to implement QEC. \textbf{B.} In Phase 2 Quantum computing, quantum computers are still confined to a single module, however the module contains enough physical qubits with low enough noise to encode logical qubits (blue). \textbf{C.} In Phase 3 Quantum computing, quantum computers grow beyond a single module, and can implement large-scale quantum algorithms fault-tolerantly.
}
\label{fig:quantum_phases}
\end{figure}

\begin{figure*}[ht]
\includegraphics[width=\linewidth]{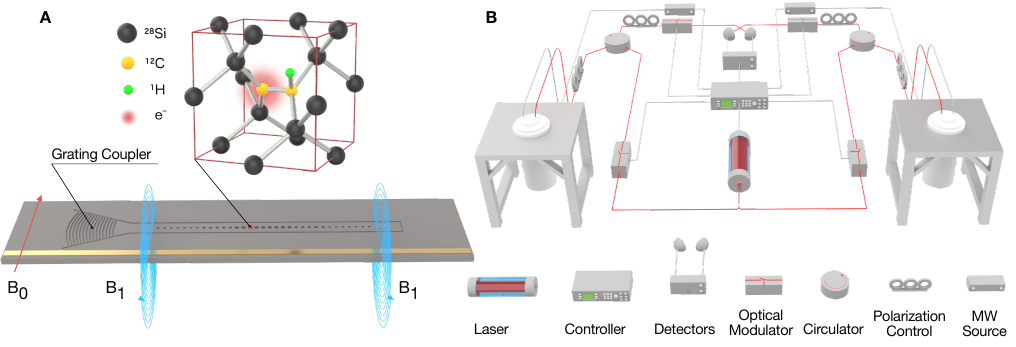}

\caption{\textbf{Schematic of the demonstration.}\label{fig:exp_schematic} 
\textbf{A.} Schematic of a T centre in the silicon lattice embedded in an optical cavity in a photonic chip. The T centre is excited with light coupled in via a grating coupler and the spin transitions are driven by microwave and radio-frequency drives delivered from a metal antenna. Magnetic field $B_0$ is applied in-plane, perpendicular to the waveguide and $B_1$ is generated from on-chip antennae in the out-of-plane direction. \textbf{B.} Schematic of the two T centre qubit modules separated by approximately 40 meters of fibre. Optical fibre is depicted in red and electrical and microwave lines in grey. The Optical Modulators are an acousto-optic modulator, electro-optic modulator, and semiconductor optical modulator on the excitation path and an acousto-optic modulator on the collection path.
}
\end{figure*}

Most quantum resource estimates for large-scale algorithms like QPE and Shor disregard the engineering requirements and constraints of a Phase 3 Quantum architecture, and assume Phase 2 Quantum monolithic design. In comparison with monolithic quantum supercomputer resource estimates, relatively little work has been devoted to networked Phase 3 Quantum supercomputer algorithm and system design \cite{Simon2016}, although there is widespread agreement that the key resource enabling networked quantum computing is entanglement, which can be consumed to perform distributed quantum operations. Remote entanglement demonstrations, so far, fall very short of the rates and fidelities that Phase 3 Quantum devices will require. At a minimum, Phase 3 Quantum computing requires distributed entanglement fidelity above the threshold of QEC codes, perhaps through entanglement distillation, at a sufficient rate to not bottleneck distributed quantum processors. For most quantum architectures under development, distributing entanglement and performing logical operations between modules will be more resource-intensive than distributing entanglement and performing logical operations within single modules.
It is therefore vital to achieve performant entanglement distribution, and to architect Phase 3 Quantum systems with the need for highly paralellized entanglement distribution in mind.

To date, the primary focus for most platforms has been improving performance within a single module, however there exist some demonstrations of remote entanglement in academic settings for many quantum computing platforms, such as: superconducting \cite{Storz2023}; quantum dots \cite{Stockill2017}; trapped ion \cite{Stephenson2020, Nadlinger2022,Ruskuc2024,Krutyanskiy2023a}; nitrogen vacancy \cite{Pompili_2021,Kalb2017,Bernien_2013}; silicon vacancy \cite{knaut2023}; and neutral atom \cite{van2022entangling}.

In this work we demonstrate key capabilities of a new quantum information platform for development towards Phase 3 Quantum applications. We present first steps toward our recently proposed distributed quantum computing and networking architecture \cite{Simmons.2023} with an initial demonstration of entanglement distribution between remote T centre quantum processor modules, and include a proof-of-principle implementation of a teleported gate sequence. 

Additionally, this paper presents the first cavity-coupled demonstrations of state initialization, single-shot nuclear spin readout, nuclear spin control, and characterization of coherence times for T centre qubits. We project rates and fidelities achievable for T centre entanglement distribution based on the Barrett-Kok (BK) entanglement protocol~\cite{Kok.2005} and best-measured T centre performance, identifying a path towards fault-tolerant distributed computation across T centre processor modules for Phase 3 Quantum computing.

\section{Demonstration of remote operations}\label{sec:demo}

 The silicon-based T centre combines a spin-photon interface with the mature nanofabrication platform of integrated silicon photonics. The T centre provides an optical interface at telecom O-band wavelength for compatibility with commercial low-loss components alongside long-lived electron and nuclear spin memories~\cite{clear2024, Xiong2024, Aberl2024, johnston2023, Simmons.2023, DeAbreu2023, Xiong2023, Lee2023, Islam2023, Sarihan2023, Simmons.2022, Higginbottom2022, Dhaliah2022,Ivanov2022,MacQuarrie_2021,Bergeron2020,Bergeron2020characterization, Costantini2006}. Integration of silicon T centres into nanophotonic devices has thus far enabled Purcell enhancement of the T centre excited state lifetime~\cite{Islam2023, johnston2023}. Integrated T centres have also shown narrow homogeneous linewidths~\cite{DeAbreu2023}, promising that the high performance bulk properties can be maintained after device integration.

Here, using the T centre platform, we demonstrate the fundamental building blocks of a distributed quantum computing or networking architecture. We benchmark the spin and optical performance of two cavity-enhanced T centres located in separate cryostats and networked via fibre optics. Isotopic purification of the silicon-on-insulator (SOI) host material enables high fidelity electron and nuclear spin initialization, control, and readout. With two high-quality spin-photon interfaces in hand, we then demonstrate the heralded generation of spin-spin entanglement between these two remote qubit systems using the BK entanglement protocol~\cite{Kok.2005, Bernien_2013}. We then consume the distributed entanglement to perform a teleported CNOT (tCNOT) gate sequence and show the truth table over a selected basis set. 

T centres were first identified as a candidate platform for quantum computing only 4 years ago~\cite{Bergeron2020}. Many improvements are anticipated in the years to come. We present projections on future performance for T centres' ability to distribute entanglement in \cref{sec:outlook}. Specifically, based upon key measured parameters, we project that T centres will be able to achieve distributed entanglement with fidelity of 0.999 for low distribution rates, or 0.998 at a rate of approximately 200 kHz.

\subsection{T Centre Characterization}\label{sec:char}

To demonstrate remote entanglement, we first prepare two silicon photonic chips with embedded T centres in separate cryostats cooled to 1.5~K and characterize the optical and spin characteristics of two remote T centres. Each cryostat is connected to an input port of a 50:50 beam splitter via 20m of optical fibre. The two T centres measured in this demonstration, labelled TC1 and TC2, are each located inside 1D photonic cavities, as shown in \cref{fig:exp_schematic}A. Optical signals are coupled via a grating coupler to an external fibre, while microwave signals for spin driving are transmitted via an on-chip antenna. The measurement apparatus is described in \cref{fig:exp_schematic}B and in the Supplementary Materials~\cref{supp_mat}.

\begin{figure}
\includegraphics[width=8.6cm]{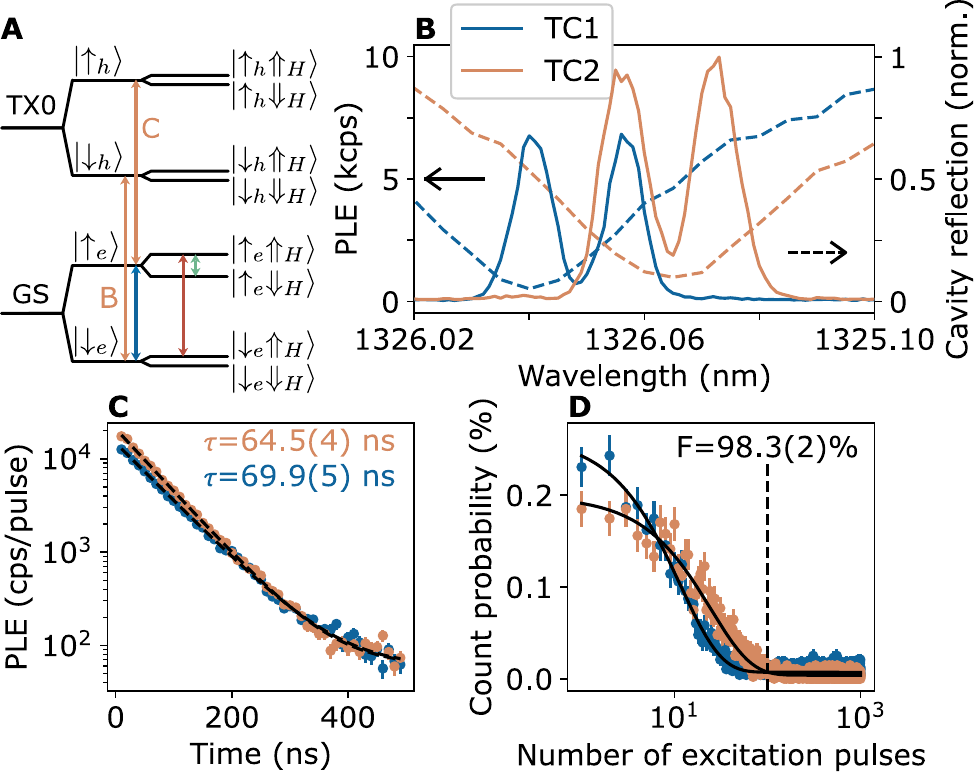}
\caption{\label{fig:exp_optical} \textbf{Individual T centre optical performance.} \textbf{A.} Fine structure of a T centre under a magnetic field showing electron ($\ket{\uparrow_e}$), excited state hole ($\ket{\uparrow_h}$), and hydrogen ($\ket{\Uparrow_H}$) spins. Splittings are (from left to right): ground state (GS) to excited state (TX0), electron/hole Zeeman splitting, and electron-nuclear hyperfine. \textbf{B.} PLE spectra (left axis) of the two T centres TC1 and TC2 at a magnetic field of $122.3$~mT, with their corresponding cavity resonances (right axis) in dashed lines. \textbf{C.} Purcell-enhanced lifetimes of the $C$ and $B$ transitions of TC1 and TC2. \textbf{D.} Initialization of the T centres from optical pumping. The dashed vertical line corresponds to the point where the initialization fidelity is $98.3(2)\%$.}
\end{figure}

In the optical ground state (GS), the T centre level structure includes electron and nuclear spin states. For simplicity, we have elected to use T centres with spin-0 carbon nuclei ($^{12}$C) for these initial demonstrations. The T centres used therefore have a single nuclear spin, provided by the hydrogen atom, as shown in \cref{fig:exp_optical}A. More generally, T centres can host up to 3 nuclear spins by exchanging each of the $^{12}$C for $^{13}$C. Optical excitation near 1326~nm resonantly generates a bound exciton at the T centre, creating the TX$_0$ state with an unpaired hole spin. In an external magnetic field, the TX$_0$ hole spin anisotropic $g$-factor and the GS electron spin isotropic $g$-factor split the optical transition into two resolved spin-selective transitions ($B$ and $C$, labeled as in Ref.~\cite{Simmons.2020}; see \cref{fig:exp_optical}A). These spin-selective optical transitions will be used for spin initialization, remote entanglement, and qubit readout.

At a magnetic field of $122.3$~mT in each cryostat, we measure (\cref{fig:exp_optical}B) the optical transitions using photoluminescence excitation (PLE) spectroscopy, and observe overlap of the $B$ and $C$ transitions of TC1 and TC2, with linewidths of $1.1$~GHz and $1.4$~GHz, respectively. In the PLE experiment we apply a microwave tone on resonance with the GS spin splitting to depopulate the dark state; this prevents state shelving under the external magnetic field. The T centres are both Purcell enhanced by the optical cavity mode with quality factors of $23,300$ \{$25,650$\}, resulting in a reduced optical lifetime from the $940$~ns bulk value~\cite{Simmons.2020} to 69.9(5)~ns \{64.5(3)~ns\} for TC1 \{TC2\}, as shown in \cref{fig:exp_optical}C. We confirm both T centres are single emitters using $g^{(2)}(\tau)$ Hanbury-Brown-Twiss correlation measurements, finding $g^{(2)}(0)=0.0076(1)$ and $0.0117(2)$, respectively (see Supplemental Materials \cref{fig:sm_g2}).

Electron spin initialization is performed by repeated cycling of the optical transition $B$ or $C$. A small probability of spin flip occurs after each cycle and the electron spin is pumped into the state opposite to that of the cycled spin-selective optical transition. \Cref{fig:exp_optical}D shows the PLE signal measured during this initialization cycle, corresponding to an initialization fidelity of 0.983(2) for TC2.

\begin{figure}
\includegraphics[width=\linewidth]{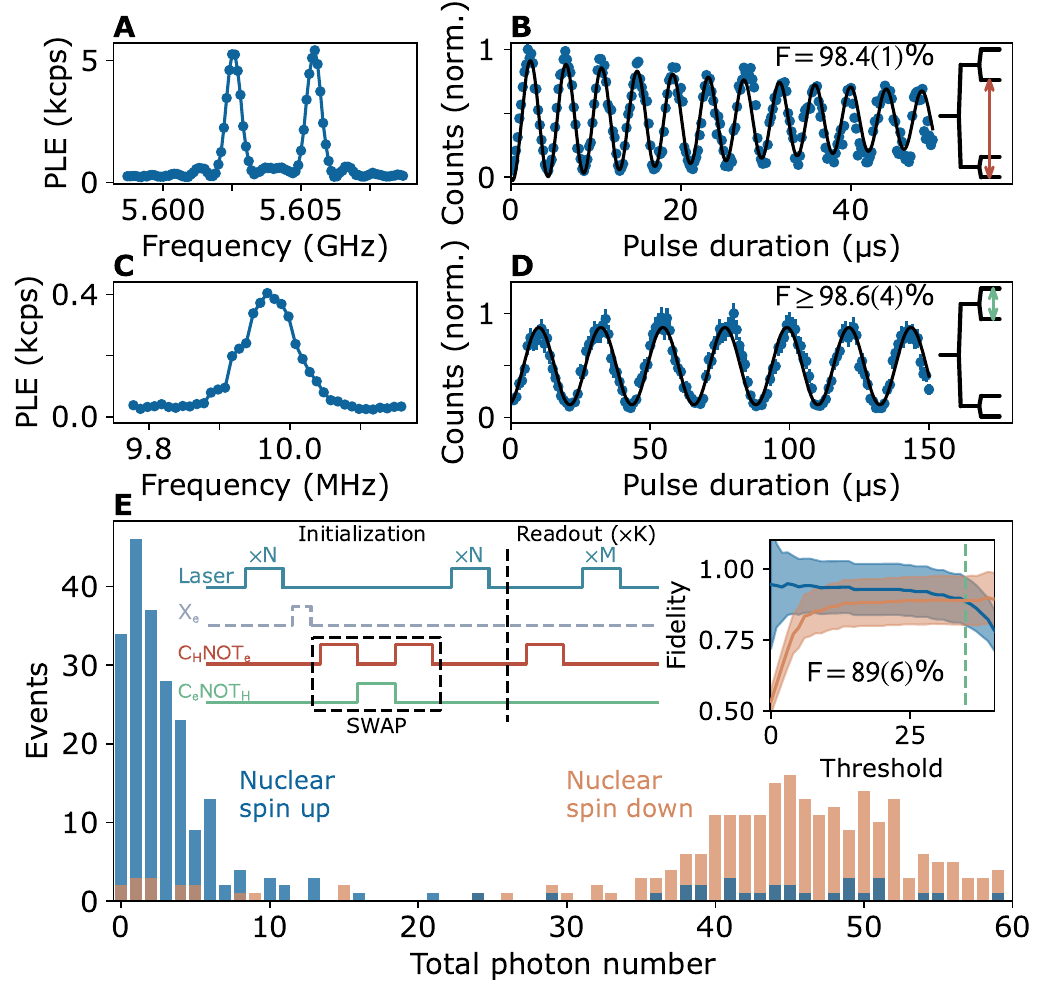}
\caption{\label{fig:exp_mw} \textbf{Individual T centre spin performance.}  \textbf{A.} ODMR showing the PLE signal as a function of MW frequency. The two resonant frequencies corresponding to nuclear-spin-selective MW transitions. \textbf{B.} Coherent Rabi oscillations driving the MW$_{\Downarrow}$ 
nuclear-spin-selective transition. A $\pi$ rotation constitutes a C$_{\rm H}$NOT$_{\rm e}$ gate. Inset: ground state energy level diagram showing the driven transition.
\textbf{C.} ODMR spectrum showing the resonant frequency corresponding to a nuclear transition. \textbf{D.} Coherent Rabi oscillations driving a electron-spin-selective nuclear transition. A $\pi$ rotation constitutes a C$_{\rm e}$NOT$_{\rm H}$. Inset: ground state energy level diagram showing the driven transition. \textbf{E.} Nuclear spin preparation and measurement: number of photons measured for a nuclear spin in the up or down state. Middle inset: the pulse sequence used to initialize and prepare the nuclear spin and readout the spin-state. Right inset: state preparation and measurement fidelity versus photon number threshold.
}
\end{figure}

We demonstrate full control of the GS spin manifold using microwave (MW) and radio-frequency (RF) controls. Optically detected magnetic resonance (ODMR) spectroscopy of nuclear-spin-selective electron transitions, MW$_\Uparrow$ and MW$_\Downarrow$, is shown in \cref{fig:exp_mw}A. Due to the hyperfine interaction, the nuclear spin-splittings are different for the two electron spin states. Rabi oscillations, driving a single transition, (\cref{fig:exp_mw}B) showcase the ability to perform a C$_{\rm H}$NOT$_{\rm e}$ gate with the hydrogen nuclear spin as the control and the electron spin as the target. 

Conversely, a C$_{\rm e}$NOT$_{\rm H}$ is possible by driving the electron-spin-selective nuclear spin transitions. \Cref{fig:exp_mw} C shows the ODMR spectrum for a swept RF drive in addition to a MW$_\Uparrow$ to generate the optical signal. A coherent Rabi oscillation is demonstrated (\cref{fig:exp_mw}D) to calibrate a C$_{\rm e}$NOT$_{\rm H}$ operation. 

Based on the decay envelope for the C$_{\rm H}$NOT$_{\rm e}$ shown in \cref{fig:exp_mw}B, we predict a gate fidelity of $98.4(1)\%$. The envelope for the C$_{\rm e}$NOT$_{\rm H}$ shown in \cref{fig:exp_mw}D does not decay sufficiently to extract a meaningful estimate of fidelity; however, by assuming that the envelope has a time constant less than $400\mu s$, we predict a lower bound of $98.6(4)\%$ on the gate fidelity. 

\Cref{fig:exp_mw}E shows the results of single shot non-demolition readout of the T centre nuclear spins after the nuclear spin has been prepared into a known initial state~\cite{jiang_2009_ssr, hesselmeier2024high, lai2024singleshot}. The pulse sequence used to initialize and prepare the nuclear spin and readout the spin-state is implemented as follows: First, the electron spin is initialized $\ket{\uparrow\Uparrow}$ (or $\ket{\uparrow\Downarrow}$ when followed by an optional non-selective electron $\pi$-pulse(X$_{\rm e}$)). An electron-nuclear SWAP gate is then performed via a C$_{\rm H}$NOT$_{\rm e}$ and a C$_{\rm e}$NOT$_{\rm H}$ gate, and finally the electron spin is re-initialized. Readout of the nuclear spin is performed by repeatedly mapping the nuclear spin to the electron spin via a C$_{\rm H}$NOT$_{\rm e}$ gate and a measurement of the electron spin with 200 low power optical pulses.
Fixing the protocol threshold at 35~photons gives a nuclear spin state preparation and measurement (SPAM) fidelity of $89(6)\%$ for TC2 as shown in the right inset of \cref{fig:exp_mw}E ($87(6)\%$ for TC1, see Supplemental Materials~\cref{fig:sm_qnd_tc1}).
T centre spin qubit performance will be further improved as we continue to develop T centres as a quantum computing platform toward the high fidelity operations demonstrated for older spin qubit systems~\cite{Bartling2024}.

\subsection{Spin coherence times}

\begin{figure*}
\includegraphics[width=\linewidth]{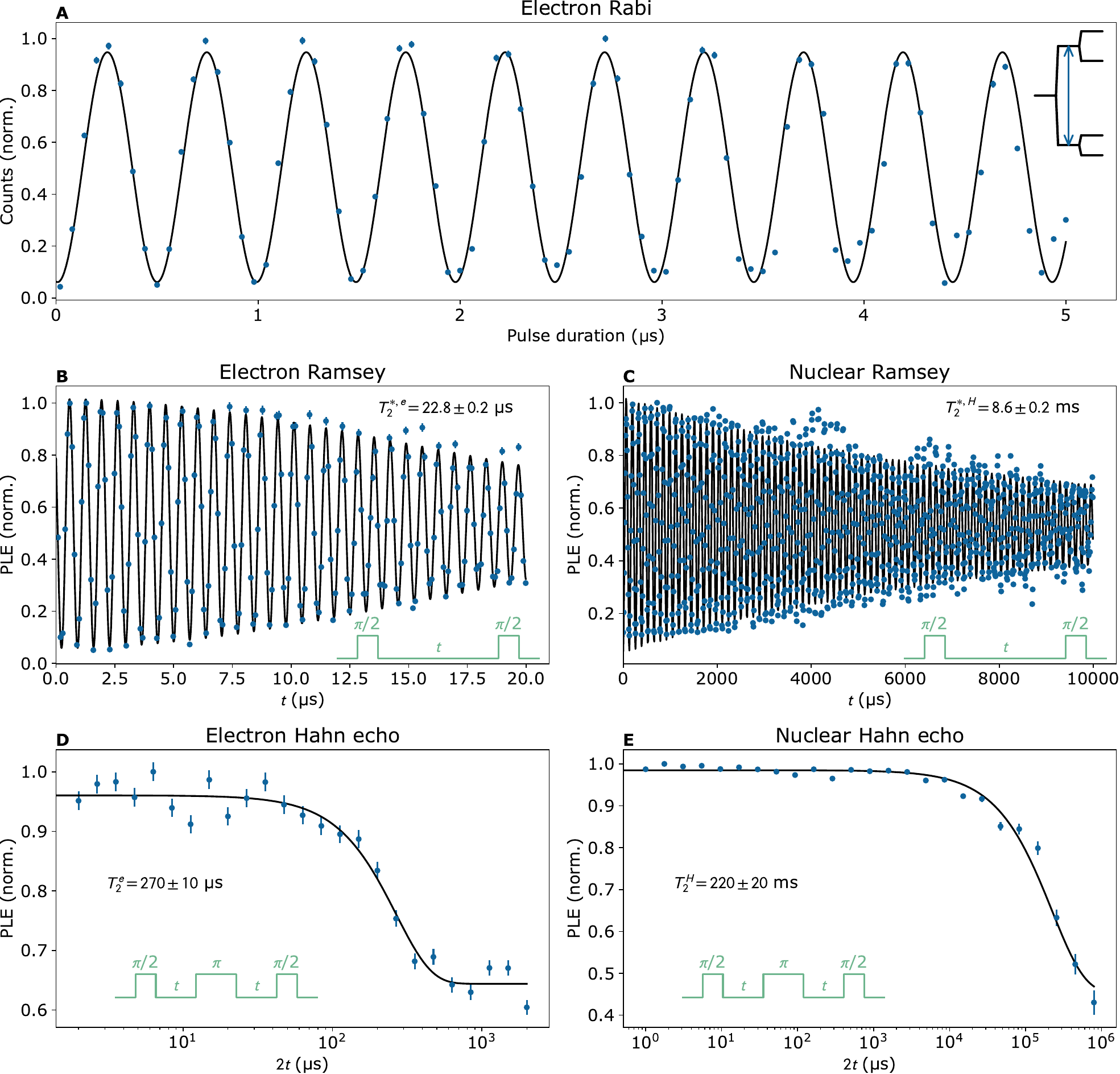}
\caption{\label{fig:sm_mw} \textbf{Spin coherence times}. \textbf{A}. Electron Rabi oscillations, driven without nuclear selectivity, the inset denotes which transition is driven. Ramsey interference fringes showing a $T_2^\ast$ decay for the electron spin, \textbf{B}, and the nuclear spin, \textbf{C}. Inset shows the pulse sequence used. Hahn-echo $T_2$ decay curves for the electron spin, \textbf{D}, and the nuclear spin, \textbf{E}. With further improvements to materials and fabrication, it is likely that the coherence times of device-integrated T centres can be improved significantly toward the coherence times observed in bulk $^{28}$Si (0.28(1) s for nuclear spins and 2.1(1) ms for electron spins~\cite{Simmons.2020}). Inset shows the pulse sequence used.}
\end{figure*}

\Cref{fig:sm_mw}A shows a Rabi oscillation of the electron spin, driven in a non-nuclear spin selective manner. Along with the CNOT pulses demonstrated in \cref{sec:char}, these results demonstrate complete control of the two spin ground state manifold. We measure the coherence times $T^\ast_2$ and $T_2$ of the electron and nuclear spins with a Ramsey pulse sequence (\cref{fig:sm_mw}B and C) and a Hahn echo pulse sequence (\cref{fig:sm_mw}D and E), respectively. For the electron we measure $T_2^\ast = 22.8(2) \upmu$s and $T_2=270(10) \upmu$s, in both cases the exponential decays are have a stretching factor of $\approx 2$ indicating residual spectral diffusion from the environment. For both electron coherence time characterizations, we drive the electron with a non-nuclear-spin-selective drive which is slightly detuned from both MW$_\Uparrow$ and MW$_\Downarrow$ transition frequencies; for the Ramsey experiment, this results in oscillations of the spin state at a frequency equal to the drive detuning. The hydrogen nuclear spin shows negligible spectral diffusion and we measure $T_2^\ast = 8.6(2)$~ms and $T_2=220(20)$~ms. The oscillations of the nuclear Ramsey is due to an intentional detuning of the drive RF field. 

Both T centres achieve spin coherence times that are the longest observed for individual solid state qubits in a commercial setting to date.
We attribute these long coherence times in part to the fact that (by design) the T centres are spatially distant from the interface;
our chips are manufactured by implanting T centres to a depth of 100 nm, placing them at a depth beyond which interfaces are not a source of reduced coherence time for other studied solid state systems \cite{Wang2016-interface}.
We therefore anticipate that T centres on similarly manufactured chips can achieve more bulk-like spin coherence times~\cite{Simmons.2020} via improved material quality and reduced system noise.

\section{Demonstration of Remote Entanglement}\label{sec:demo}

Two-photon interference is a key mechanism for generating photon entanglement, and the Hong-Ou-Mandel (HOM) experiment is a well-known demonstration of this effect \cite{HongOuMandel_1987}. In the experiment, two photons are input into separate ports of a beam splitter, interfere with each other, and are measured by detectors at each output of the beam splitter. Coincidence counts are measured as a function of time difference between detector clicks. For perfectly indistinguishable photons, these counts drop to zero for all time differences, signifying the photon bunching effect which indicates quantum interference.

The BK protocol is an emission-based strategy that can be used to generate entanglement between qubits \cite{Kok.2005}. The protocol is as follows: the qubits to be entangled are each prepared in the $\ket{+}=\frac{1}{\sqrt{2}}(\ket{\uparrow}+\ket{\downarrow})$ state and the $\ket{\uparrow}$ state is optically cycled to entangle the presence of a photon with the spin state. The emitted photons are directed to a beamsplitter with detectors at the output ports, as in the HOM experiment. Next, the $\ket{\uparrow}$ and $\ket{\downarrow}$ states are exchanged via a MW $\pi$-pulse and the emission and detection are repeated. A single detection event for both rounds heralds the successful projection of the spins into a Bell pair (BP) with the phase of the BP determined by the specific pattern of detector clicks. In the case of T centres, we implement this protocol to generate distributed entanglement between the electron spins in different T centres.

In emission-based entanglement strategies, such as the BK protocol, the photons emitted from distributed qubits must be highly indistinguishable~\cite{Bernien_2013}. The degree of indistinguishability between two photon sources can be quantified by their HOM interference visibility, $V$. Notably, when two photons are distinguishable via a slight frequency mismatch, the high timing resolution of single photon detectors can be used to apply a ``time-bin filter'' to the emitted photons~\cite{metz2008, kambs2018}. By selecting only photon pairs detected with a narrow time difference $d\tau$, the effective visibility remains high even for imperfect frequency matching. For solid state single photon sources, noise in the local environment can cause spectral diffusion of the optical transitions and introduce an uncertainty in the frequency of the emitted photons. Time-bin filtering therefore allows the emission of solid state sources to maintain high indistinguishability even in the presence of this uncertainty. 

\begin{figure}
\includegraphics[width=8.6cm]{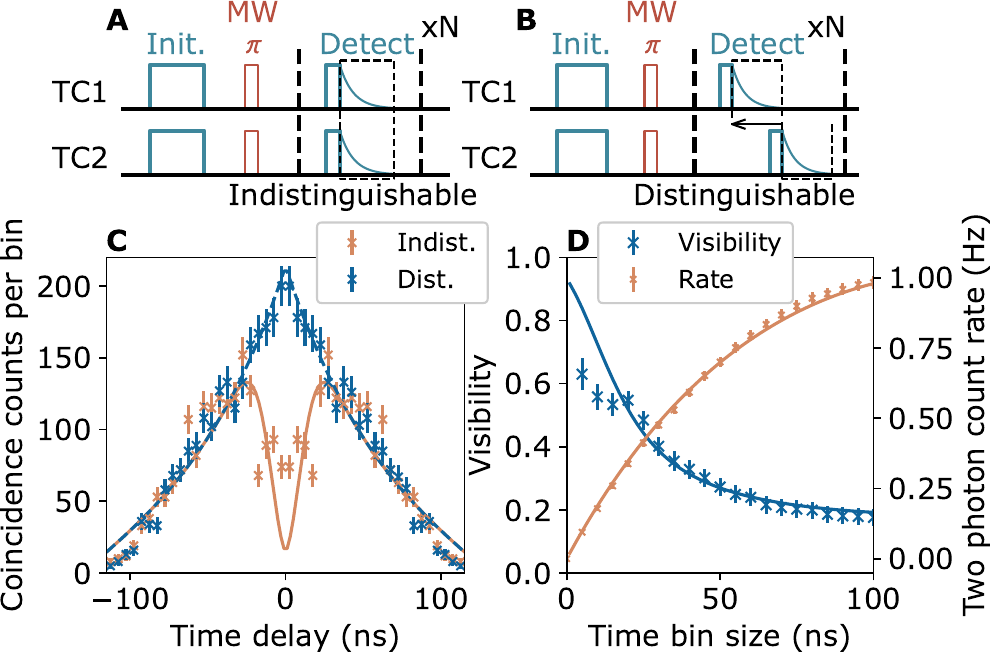}
\caption{\label{fig:HOM} \textbf{Hong-Ou-Mandel two-photon interference.} \textbf{A.} Pulse sequence used to collect indistinguishable events. \textbf{B.} Pulse sequence used to collect distinguishable events. In the analysis, time tags from the delayed optical pulses of TC2 are shifted back to align with the first optical excitation and detection of TC1, as indicated by the leftward arrow. \textbf{C.} Measured coincidence counts for distinguishable (orange cross) and indistinguishable events (blue cross) with Poissonian error. Models using second-order correlation functions are overlaid in solid blue and dashed orange, respectively. \textbf{D.} Two-photon visibility as a function of time filter size (blue cross) from the data in A, shown with model (blue dashed line). Distinguishable coincidence count rate as a function of filtering (orange cross) with Poissonian error and modelled rate (orange solid line).}   
\end{figure}

We obtain the visibility $V$ of photons emitted from TC1 and TC2 by measuring the coincidence clicks from indistinguishable and distinguishable photons using the sequence in \cref{fig:HOM} A and B (see details in the Supplementary Material \cref{supp_mat}). By time-tagging the photon events according to their time difference, we show in \cref{fig:HOM}D the visibility as a function of the time-bin filter size. For a small time-bin of $5$~ns, photons are indistinguishable with $V=0.63$, however this comes at the price of a two-photon rate equal to $0.09$~Hz. Additional data in the Supplementary Materials \cref{fig:sm_low_power_hom} shows a maximum of $V=0.87$ at time bin size of $10$ns, with rate of $2$~mHz. Inversely, for a large time-bin of $100$~ns, the rate climbs to $1$~Hz with $V\approx0.2$. The model, for which parameters are included in Supplementary Materials \cref{tab:Measured performance parameters}, and data shows good overall agreement. However, below a time bin of $20$~ns we see a saturation of the measured visibility or, equivalently, the coincidence plot does not dip to zero. We expect this is due to current limitations in our measurement hardware, and anticipate that future development will lead to improved performance. In \cref{sec:outlook}, we quantify performance projections for distributed entanglement.

For entanglement generated by the BK protocol, the BP fidelity can be upper-bounded using the HOM visibility as $F\leqslant(1+V)/2$~\cite{Bernien_2013}. This maximum fidelity can be considered as an upper bound on the entanglement fidelity as we neglect here other causes of infidelity from local gates, finite branching ratios and double excitation probabilities. The BP generation rate can be upper bounded at half the HOM rate since the BK protocol includes two repetitions of the sequence used in a HOM experiment. This estimate neglects the initialization and microwave portions of the BK protocol. From the results of this particular HOM demonstration, we project a HOM-based upper bound of $F=0.96$ and a rate of $7.5$~mHz at $1$~ns time bin size.

\begin{figure}
\includegraphics[width=8.6cm]{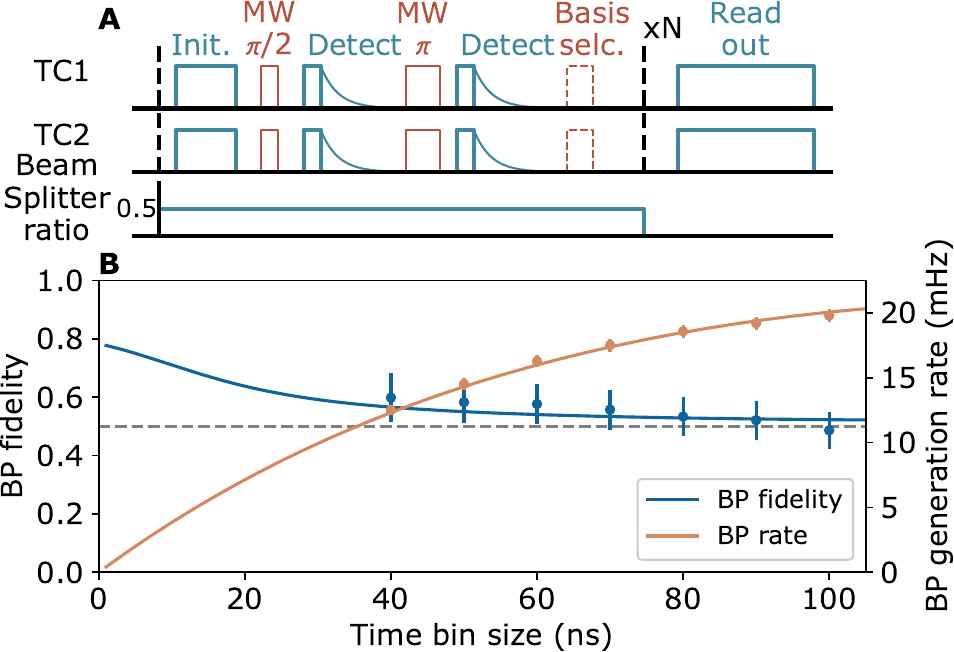}
\caption{\label{fig:BK_binning} \textbf{Demonstration of the Barrett-Kok protocol.} \textbf{A.} Pulse sequence for BP entanglement. We initialise the two T centres by optical pumping and then perform the entangling sequence. The final microwave pulse determines the basis selection for readout. \textbf{B.} BP generation rate and fidelity for different time bin size.}
\end{figure}

Harnessing the indistinguishable T centre photon emission, we generate entanglement by applying the BK protocol~\cite{Kok.2005}. A breakdown of the protocol's pulse sequence used is shown in \cref{fig:BK_binning}A with an early-late time bin separation of 1910~ns. Fast feedback triggers a readout when both an early and late photon are measured. We confirm the presence of entanglement by applying MW gates (dashed pulse in \cref{fig:BK_binning}) to select different readout bases (see \cref{sub_sec:BKprotocol}).

\Cref{fig:BK_binning}B shows the measured BP entanglement fidelity reaching $F=0.60(0.08)$ for a $40$~ns time bin size. Both the BP fidelity and BP generation rate match the simulated results, taking into account optical loss, excitation probability, known T centre optical properties, and local gate errors (see Supplementary Materials \cref{tab:Measured performance parameters} for more details).
We find that for time bin widths less than $60$~ns, the measured fidelity with one sigma error bars lies above the threshold to witness BP entanglement $>50\%$ \cite{Sackett2000}.
The performance does not meet the projected bound based on the HOM protocol because the experiment includes sources of infidelity not considered in the calculation of the bound such as initialization, measurement, and local operations, etc.
Higher photon detection rates would allow for more selective time bin filtering, which we expect would produce a BP fidelity of 80$\%$.

\begin{figure}[t!]
    \includegraphics[width=\linewidth]{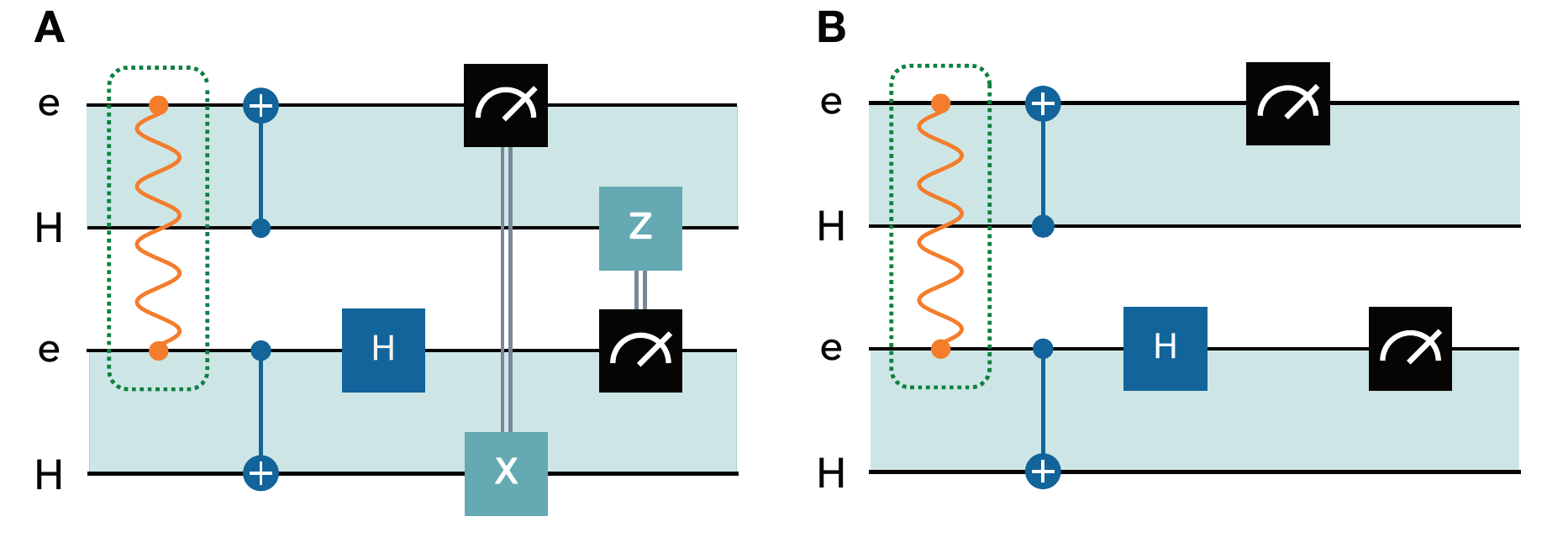}
    \caption{\textbf{Teleported CNOT circuit diagrams.} \textbf{A}. A tCNOT implemented between Hydrogen nuclei (H) in T centres in different modules. We highlight the space between the electron (e) and nuclei in the same T centre for illustration purposes. The first step (in a dashed box) is to establish a distributed Bell pair on the electrons of the two T centres. Next, we implement local measurements and operations in each T centre to complete the tCNOT. \textbf{B}. Post-selected teleported CNOT circuit. A post-selected tCNOT can be implemented by omitting the feed-forward operations and post-selecting on measurement outcome 00.
    }
    \label{fig:tCNOTcircuit}
\end{figure}

\begin{figure}[t!]
\includegraphics[width=8.6cm]{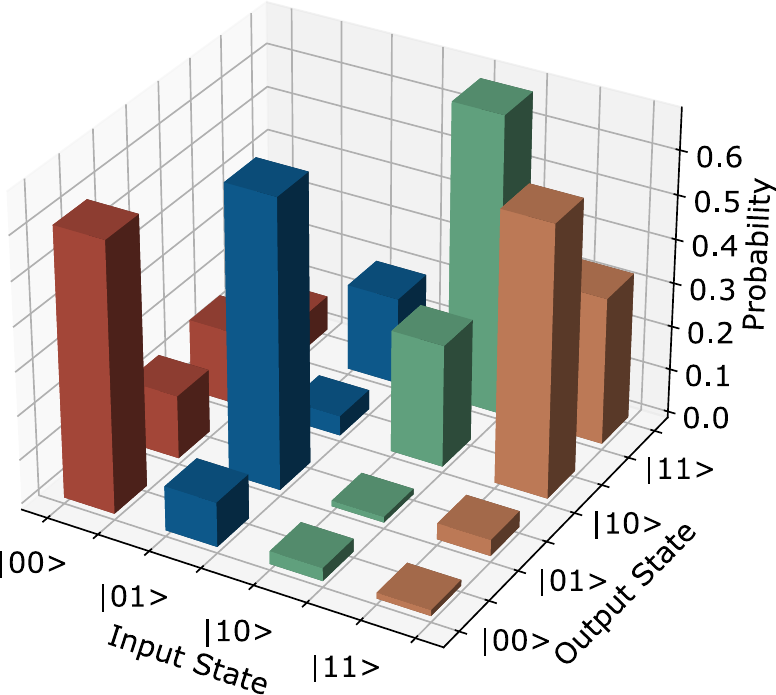}
\caption{\label{fig:tCNOT} \textbf{Truth table for preliminary  T centre teleported CNOT sequence}. Truth table for initial tCNOT experiment, using post-selected CNOT.
}
\end{figure}

Armed with distributed BP's, we demonstrate the ability to perform a tCNOT gate sequence to assess the initial capabilities of the system~\cite{Eisert_2000}. A tCNOT sequence is described in \cref{fig:tCNOTcircuit}A and is composed of entanglement generation via the BK protocol, local operations, single qubit measurements, and feed forward operations. In this experiment, we forego all feed forward gate operations. Instead, we perform the same tCNOT sequence by post-selecting the successful photon detection case from the electron-spin readout and generate the tCNOT truth table for this experiment (shown in \cref{fig:tCNOT}).
The characterization for \cref{fig:tCNOT} was implemented by prepending state preparation and appending nuclear spin measurements to the circuit in \cref{fig:tCNOTcircuit}B. \footnote{The state preparation is implemented by preparing the electron state, performing a SWAP$_{eH}$ operation, and then re-initializing the electron state.}
For an unrestricted input space, this modified sequence only executes a tCNOT for one pattern of BK and measurement outcomes.
 The local C$_{\rm e}$NOT$_{\rm H}$ and C$_{\rm H}$NOT$_{\rm e}$ operations are conditionally applied when the requisite BK outcomes are observed.
 This sequence demonstrates a method for remote gate operations in a distributed quantum processor, key to the development of Phase 3 Quantum information technologies.

\section{Outlook}\label{sec:outlook}

Having established T centres as one of only a few systems to achieve remote entanglement distributed between two separate modules over a telecom fibre network, we now analyze how further device-level integration and development of the T centre devices will build upon this demonstration, with corresponding improvements in both rate and fidelity to enable effective distributed quantum computing for Phase 3 Quantum technology.

Performance  improvements can be divided into two categories: integration and material development. We will assume integration into a photonic and fibre network with best-in-class components, and T centres with the best-measured spectral properties. These are not fundamental limits, but merely the best measured spectral properties in the four years T centres have been studied as spin-photon interfaces for quantum technologies. T centres measured in bulk samples of isotopically pure, single-crystal silicon have demonstrated intrinsic optical dephasing rates of $230$~kHz \cite{DeAbreuThesis2024}, and slow spectral diffusion of $20$~MHz \cite{MacQuarrie_2021}. We will take these bulk T centre properties as the eventual performance of integrated centres, although improvement beyond this level remains possible. Further to this spectral performance, we will assume an increase in the cavity Purcell factor over the current designs presented in \cref{sec:demo} by a combination of higher $Q$ and reduced mode volume to give an emission lifetime of $\tau = 10$~ns.

\begin{figure}[t!]
\includegraphics[width=8.6cm]{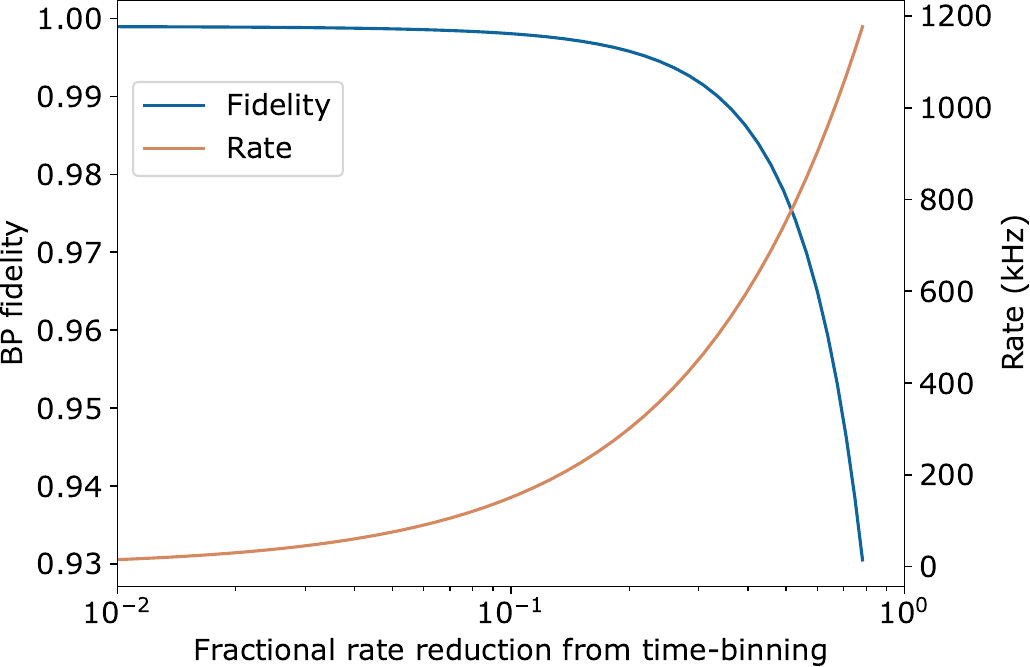}
\caption{\label{fig:ceiling_rate_fidelity} \textbf{Fidelity and rate of remote Bell pair distribution using projected T centre performance}. A time-window correlation filter is chosen which increases entanglement fidelity and captures a fraction of the total coincidences. Additional sources of infidelity outside of the emission spectral coherence are considered including double excitation and off-resonant excitation.}
\end{figure}

We measure performance in terms of achievable entanglement attempt rates, success probabilities, and the fidelity-success probability trade-off curve. We calculate a feasible repetition rate of $3.4$~MHz from the times for each step of the BK entanglement scheme in \cref{tab:ceilingRepetitionRates}, including signal latencies and feedback times to correct the state based on detection patterns. The expected success probability is $44$\%, calculated by compounding the effective loss mechanisms shown in \cref{tab:ceilingSuccessRates}, consisting of intrinsic `loss' mechanisms such as the BK success probability, optical path losses between remotely coupled photonic chips, and T centre performance, but not including a time-filter. Accounting for the intrinsic success probability of the BK protocol, the maximum possible success for a perfect system is $50$\%, meaning that we project that T centres can achieve $88$\% of the intrinsic maximum.

The fidelity of the entangled state is calculated considering the fidelity-rate trade-off curve of the BK protocol with interference visibility limited by the emitters' spectral purity following Ref.~\cite{kambs2018}. The optimal time-filter window depends on the additional sources of infidelity. We consider additional infidelity due to the probability of a second excitation during an $83$~ps excitation pulse chosen to cause negligible off-resonant excitation at a field of $500$~mT.

Under these constraints, the maximum attainable fidelity is $F=0.999$ in the low-rate limit. A compromise time window that collects 12.5\% of total coincidences yields $F=0.998$, with an entanglement rate of approximately $200$~kHz. In \cref{fig:ceiling_rate_fidelity} we show the rates and fidelities as a function of the total coincidence fraction that is filtered. At reasonable rates these fidelities surpass the thresholds required for both entanglement distillation and QEC codes. These forecasts are by no means the performance ceiling for remote entanglement with T centres. Additional material improvements and new control techniques (see e.g. \cite{Schweickert2018, Coste2023}) could all increase the attainable rate and fidelity of entanglement distribution.

\begin{table}[!t]
\centering
\small
\begin{tabularx}{\columnwidth}{
  >{\raggedright\arraybackslash\hsize=1\hsize}X
  >{\centering\arraybackslash\hsize=1\hsize}X
  >{\centering\arraybackslash\hsize=1\hsize}X
  >{\centering\arraybackslash\hsize=1\hsize}X
}
\toprule
Step & Time (ns) & Number & Total time (ns) \\
\midrule
\rowcolor{lightgray}
Initialization & 100 & 1 & 100\\
Excitation & 1 & 2 & 2\\
\rowcolor{lightgray}
Detection & 30 & 2 & 60\\
Pi pulse & 10 & 1 & 10\\
\rowcolor{lightgray}
Signal latency & 10 & 2 & 20\\
Feedback & 100 & 1 & 100\\
\midrule
\textbf{Total} & & & \textbf{292}\\
\bottomrule
\end{tabularx}
\caption{Minimum times for each step in the BK protocol, leading to a total minimum success time of $292$ ns, or a maximum repetition rate of $3.4$ MHz. }
\label{tab:ceilingRepetitionRates}
\end{table}

\begin{table}[!t]
\centering
\small
\begin{tabularx}{\columnwidth}{
  >{\raggedright\arraybackslash\hsize=1\hsize}X
  >{\centering\arraybackslash\hsize=1\hsize}X
  >{\centering\arraybackslash\hsize=1\hsize}X
  >{\centering\arraybackslash\hsize=1\hsize}X
}
\toprule
Element & Efficiency & Number & Total efficiency \\
\midrule
\rowcolor{lightgray}
BK intrinsic & 0.5 & 1 & 0.50\\
Excitation & 0.99 & 2 & 0.98\\
\rowcolor{lightgray}
Emission & 0.99 & 2 & 0.98\\
Detector & 0.99 & 2 & 0.98\\
\rowcolor{lightgray}
Chip-fibre & 0.97 \cite{psiQhw} & 2 & 0.94\\
\midrule
\textbf{Total} & & & \textbf{0.443}\\
\bottomrule
\end{tabularx}
\caption{Projected success rates for each step in the BK protocol, leading to a total success probability of $44.3\%$. Note that in a perfect system the BK protocol achieves a 50\% success rate.
}
\label{tab:ceilingSuccessRates}
\end{table}

\section{Conclusion}

Quantum computers can scale to the level required for commercially relevant algorithms with a modular architecture. High rate, high fidelity distributed entanglement will be necessary to implement algorithms on such an architecture effectively and, without careful consideration, can dominate the resource requirements for a distributed architecture \cite{Simon2016}. 

In this paper, we presented a first demonstration of distributed entanglement between T centres, one of only a handful of colour centres to achieve remote entanglement. These entangled qubits are on silicon photonic chips, each capable of hosting and controlling thousands of qubits, in separate cryostats and connected by optical fibre. These qubits can be connected via optical fibres, making their operation compatible with optical fibre switch networks, and allowing the system to be extended horizontally to many more modules with high connectivity. The T centre emits in a telecommunications band; this demonstration can therefore be performed over tens or hundreds of kilometres without frequency conversion. This band, the telecom O-band, is emerging as a popular quantum network standard \cite{Gotham2024, Makris2023, Schrenk2019} and potentially the future ``quantum band".

This entanglement demonstration between remote processors establishes a fundamental building block for a scalable Phase 3 Quantum computer, establishing T centres as a candidate Phase 3 Quantum architecture.
Cross-module operations will be essential for the execution of distributed, fault-tolerant quantum algorithms in commercially valuable Phase 3 Quantum computing. Finally, we considered entanglement distribution performance prospects for future T centre devices using the same Barett-Kok entanglement scheme but with a higher degree of chip integration and the best-measured T centre material properties. The achievable rates and fidelities far exceed previous optically distributed entanglement demonstrations \cite{Stockill2017,Stephenson2020, Nadlinger2022,Ruskuc2024,Krutyanskiy2023a,Pompili_2021,Kalb2017,Bernien_2013,knaut2023,van2022entangling}. This result unlocks distributed quantum computing in silicon, and a path towards networks of silicon quantum processors performing commercially and socially transformative calculations.

\bibliography{references}

\section{Supplementary Material}\label{supp_mat}

\subsection*{Experimental Setup}
\label{sub_sec:apparatus}
The chips are made from silicon-on-insulator wafers; the device layer is 200~nm thick silicon that hosts the T centres. The T centres are created by a four-step process developed by Ref.~\cite{MacQuarrie_2021}.
Each quantum chip is mounted on a copper mount on a three-axis stage in a closed-cycle cryostat at 1.5~K. The stage allows for precise alignment between the on-chip grating couplers and a fibre array. The fibre bundle connected to the array is passed through a vacuum feedthrough to room-temperature for connection to the control equipment. Radio and microwave frequencies are transmitted to the on-chip antennas via wirebonds to a printed circuit board (PCB), from where cables are similarly routed to room-temperature.

The optical setup as illustrated in \cref{fig:exp_schematic}B is based on pulsed resonant excitation followed by time-filtering of the T centre photoluminescence. For each setup, a tunable O-band laser provides the excitation at the T centre wavelength ($\approx$~1326~nm), stabilized by feedback with a wavemeter, and gated by a semiconductor optical modulator, an acousto-optic modulator, and an electro-optic modulator in series. The combination of modulators allow for nanosecond pulsing with $\gg$~80~dB intensity extinction ratio. Electronic optical variable attenuators provide tuning of the optical power. The light is coupled in and out of the fibre array by an optical circulator and a polarization controller. On the collection side, an AOM gates the collection to prevent latching and long deadtime of the superconducting nanowire single photon detectors. The photons from each chip are passed through a variable beam-splitter before the detectors, allowing either individual readout or correlation measurements. The detector clicks are binned and/or time-tagged using a time-tagger and further analyzed on a computer.

Pulses for nanosecond timing control are generated using an arbitrary waveform generator, and is used for microwave excitation via an IQ modulator at 3.424~GHz and a microwave amplifier.

\FloatBarrier

\subsection*{Further T Centre Characterization}

\begin{figure}
\includegraphics[width=8.6cm]{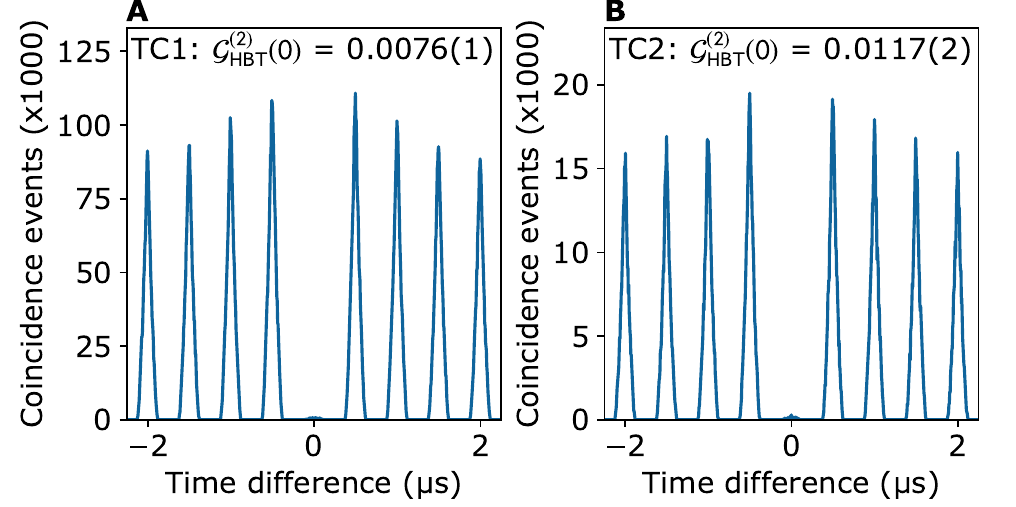}
\caption{\label{fig:sm_g2} \textbf{Hanbury-Brown-Twiss correlation measurement}. Coincidence measurements of the output of a HBT experiment beamsplitter versus time difference for TC1, \textbf{A}, and TC2, \textbf{B}.}
\end{figure}

For the visibility and fidelity calculations the two photon component of the T centre emission is required. To this end a Hanbury-Brown-Twiss measurement is performed to measure the HBT second-order correlation. We excite each T centre individually, route the photons to a beamsplitter, and measure the coincidence clicks of the two detectors on the beamsplitter output. The T centres are pulsed with a 500~ns repetition rate. To determine $\mathcal{G}_{HBT}^{(2)}$ we normalize the coincidence clicks by the integrated area of the outer most peaks. The results for both TC1 and TC2 are shown in \cref{fig:sm_g2} showing $\mathcal{G}_{HBT}^{(2)}(0) = 0.0076(1)$ and $0.0117(2)$ respectively.

\begin{figure}
\includegraphics[width=8.6cm]{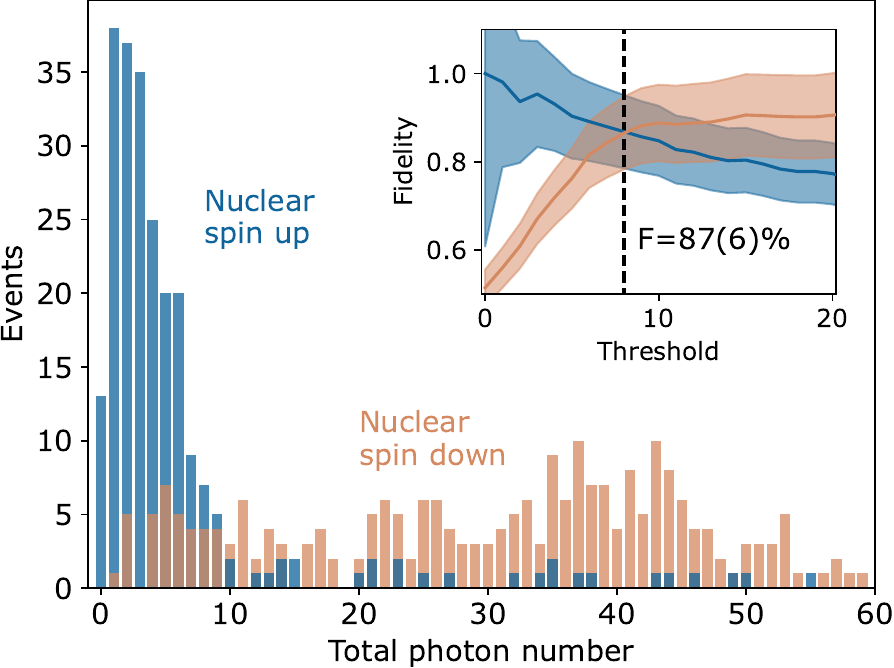}
\caption{\label{fig:sm_qnd_tc1} \textbf{TC1 single-shot readout.} Nuclear spin preparation and measurement: number of photons measured for a nuclear spin in the up or down state. Right inset: state preparation and measurement fidelity versus photon number threshold.}
\end{figure}

Single shot readout of TC1 is demonstrated in \cref{fig:sm_qnd_tc1}, measured in the same manner as shown for TC2 in \cref{fig:exp_mw} of the main text. In this case the state preparation and measurement fidelity is 87(6)$\%$ at a threshold of 8 photons. 

\subsection*{Two photon interference}
\label{sub_sec:tpi}

To measure the HOM visibility we begin by initializing the two T centres by repeated optical excitation of the two resonant transition: B and C of TC1 and TC2 respectively. A MW pi-pulse is applied to flip the spins to be back into resonance with the laser. The two T centres are excited with a low power optical pulse to have the emission from each T centre to arrive at the beamsplitter at the same time in order to generate counts in the indistinguishable case. This step is repeated until the system needs to be re-initialized. After the re-initialization, the distinguishable case is measured by delaying the optical excitation pulse of TC2 by 500 ns so the two photons arrive at different times. This entire pulse sequence is repeated many times to accrue sufficient statistics.

For the analysis the timetags from the delayed optical pulses of TC2 are shifted back by 500 ns to align with the first optical excitation and detection of TC1. Correlations from the two detectors at the output of the beamsplitter are then determined for both the indistinguishable and distinguishable cases.

We measure both distinguishable $(D)$ and indistinguishable $(I)$ coincidence counts to find the two-photon visibility defined by
\begin{equation}\label{eq:Vis}
    V=1-\frac{P^{I}_{\text{coinc}}}{P^{D}_{\text{coinc}}},
\end{equation}
where $P^{I}_{\text{coinc}}/P^{D}_{\text{coinc}}$ gives integrated area of second order cross-correlation function $\mathcal{G}_{HOM,I(D)}^{(2)}(\tau)$ for the corresponding measurement.
For an ideal experimental set up with perfect spacial overlap and 50:50 beam splitters $V \equiv \mathcal{I}$. For the measured and simulated (in)distinguishable coincidence counts see \cref{fig:HOM}(a). To model (in)distinguishable cross-correlation functions we use the general expression from \cite{kambs2018} supplement. We also account for non-perfect single photon emission captured with each emitter's Hanbury Brown and Twiss second order correlation function ${\mathcal{G} }_{HBT}^{(2)}(\tau)$, with $\mathcal{G}_{HBT_{T1}}^{(2)}(0) = 0.008\pm 0.006$ and $\mathcal{G}_{HBT_{T2}}^{(2)}(0) = 0.012\pm 0.010$ by \cite{Sipahigil_2012}, 
\begin{equation}\label{eq:HOM}
    \mathcal{G}_{HOM,_{I(D)}}^{(2)}(\tau) = \frac{1}{2}\mathcal{G}_{I(D)}^{(2)}(\tau) + \frac{1}{4}\big(\mathcal{G}_{HBT_{T1}}^{(2)}(\tau) + \mathcal{G}_{HBT_{T2}}^{(2)}(\tau)\big).
\end{equation}
Where $\mathcal{G}_{I}^{(2)}(\tau) =  \mathcal{G}^{(2)}_0(\tau) + \mathcal{G}^{(2)}_{int}(\tau)$ 
 is the second order correlation function for the indistinguishable case and $\mathcal{G}^{(2)}(\tau) =  \mathcal{G}^{(2)}_0(\tau)$ is for perfectly distinguishable photons as defined in \cite{kambs2018}. Noting that the normalisation used in (\cref{eq:HOM}) to combine the HBT correlation functions assumes perfect beamsplitter reflection/transmission. 

Parameters used for our model of the visibility are $5$MHz homogeneous broadening \cite{DeAbreu2023}, $22.5$MHz excitation bandwidth which sets the range of inhomogeneous broadening (if the emitter frequency has spectrally wandered outside this bandwidth it is solely contributes to loss and effects the photon rate only). To account for experimental imperfections we include an estimated polarisation mismatch of $12.8^\circ$.

\begin{figure}

\includegraphics[width=8.6cm]{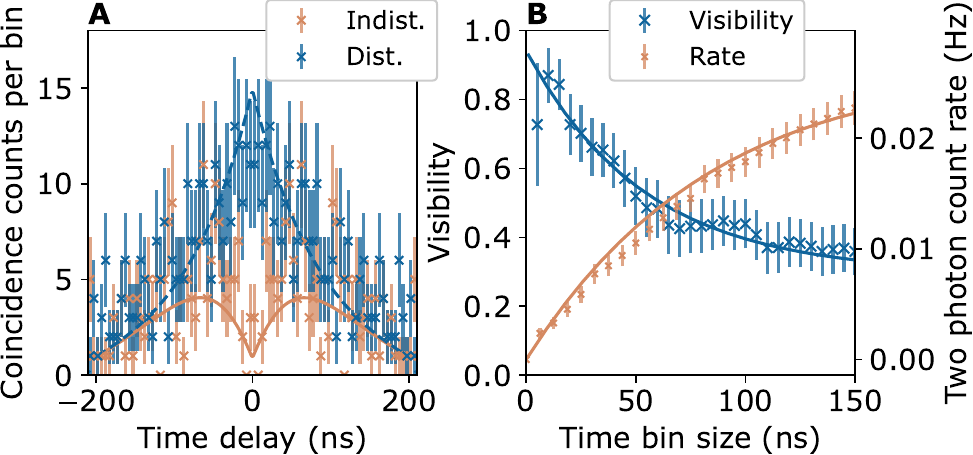}
\caption{\label{fig:sm_low_power_hom}\textbf{Low-power HOM}. \textbf{A.} (In)distinguishable coincidence counts with Poissonian error bars for two T centres (TC1 and TC2) with low power laser (estimated excitation bandwidth is $3$MHz and homogeneous linewidths of $4$MHz). The lifetime for these emitters were measured at $100(2)$ns and $77(1)$ns, and the acquisition window was $240$ns. \textbf{B.} Corresponding visibility and distinguishable coincidence count rate. Model uses the same experimental imperfections listed in \cref{tab:Measured performance parameters} and shows Poissonian error.
}
\end{figure}

The model of the two photon count rate is found by considering the joint average excitation and detection probability and the two photon time binning loss from $\eta_{tb} = \int^{\tau_B}_{-\tau_B} g_{HOM_D}^{(2)}(\tau)/ \int^{\tau_{lim}}_{-\tau_{lim}} g_{HOM_D}^{(2)}(\tau)$. To account for the reduction of counts at the tails we adapt the work in ref. \cite{kambs2018} to include a truncation of the absolute time to $\tau_{lim}=130$ns to capture finite acquisition window. The HOM measurement was repeated at a lower power and correspondingly lower rates (see \cref{fig:sm_low_power_hom}) to yield a higher fidelity. 

\FloatBarrier

\subsection*{Barrett-Kok protocol}
\label{sub_sec:BKprotocol}
To determine the fidelity of the generated BP~\cite{Jozsa1994} we measure in either the population ($\ket{01},\ket{00}$) or the coherence basis ($\ket{++},\ket{+-}$) after a successful round of entanglement. To readout in the population basis we apply microwave pulses to the two electron spins to rotate the spins by $\{0,0\}$ for $\ket{00}$ or $\{0,\pi\}$ for $\ket{01}$. For the coherence basis we rotated the spins to a superposition state: we apply $\{\pi/2,\pi/2\}$ to readout $\ket{++}$ and $\{\pi/2,-\pi/2\}$ for $\ket{+-}$.

We define the BP fidelity by combining the population and coherence basis measurements as $F=(P+C)/2$ \cite{Wang2016}. With $P=N_{01}/(N_{00}+N_{01})$ and $C=(N_{++}-N_{+-})/(N_{++}+N_{+-})$ where $N$ are the binned counts for the corresponding measured state. We bin the counts based on the arrival time of the late photon relative to the early photon with the time difference of the early/late excitation pulses (1910~ns) subtracted.

To model the generated entangled state we construct an effective quantum channel composed of non-unitary quantum channels corresponding to each of the operations in the BK protocol. These constituent channels account for two-photon indistinguishability; finite (non)radiative branching ratio; infidelity of electron spin rotations and initialization; false heralding from dark counts; and double photon emission probability~\cite{Humphreys2018}. For parameters used see \cref{tab:Measured performance parameters}. For the radiative branching ratio (BR) we measure for each T centre the spin-photon correlation by taking the fraction between the late and early clicks with the BK sequence in both coherence and population measurement basis. These values have been found to be $4\%$ and $6\%$. We estimate the non radiative BR is equal to the radiative BR at $2.5\%$ to best match the measured BP fidelity.
The rate for BP generation has been found by multiplying together the total photon loss, time bin loss, spin projection loss $(50\%)$ and the repetition rate.

\begin{table}[!t]
\centering
\small
\begin{tabularx}{\columnwidth}{
  >{\raggedright\arraybackslash\hsize=1.9\hsize}X
  >{\centering\arraybackslash\hsize=0.7\hsize}X
  >{\centering\arraybackslash\hsize=0.7\hsize}X
  >{\centering\arraybackslash\hsize=0.7\hsize}X
}
\toprule
Parameter & TC1 & TC2 & unit \\
\midrule
\rowcolor{lightgray}
Purcell-enhanced lifetime ($\tau$) & 69.9(5) & 64.5(4) & ns \\
Fast spectral diffusion ($\Gamma*$) & \multicolumn{2}{c}{5(2)$^\star$} & MHz \\
\rowcolor{lightgray}
Excitation bandwidth ($\sigma$); captures range of slow spectral diffusion & \multicolumn{2}{c}{22.5(3.0)$^\star$} & MHz \\
Polarisation mismatch & \multicolumn{2}{c}{12.8$^\star$} & $^\circ $\\ 
\rowcolor{lightgray}
Photon arrival time desynchronisation & \multicolumn{2}{c}{0(5)$^\star$} & ns\\
Darkcount rate & \multicolumn{2}{c}{10} & Hz \\
\rowcolor{lightgray}
Average gate fidelity & \multicolumn{2}{c}{98.575$^\star$} & \% \\
Acquisition window & \multicolumn{2}{c}{130} & ns \\
\rowcolor{lightgray}
Radiative branching ratio & \multicolumn{2}{c}{2.5$^\star$} & \% \\
Non-radiative branching ratio & \multicolumn{2}{c}{2.5$^\star$} & \% \\
Detector loss & \multicolumn{2}{c}{-1.97} & dB \\
\rowcolor{lightgray}
Symmetric cavity loss & \multicolumn{2}{c}{-3} & dB \\
Path loss & \multicolumn{2}{c}{-7} & dB \\
\rowcolor{lightgray}
Probability of excitation &  \multicolumn{2}{c}{-14.9} & dB \\
Quantum efficiency loss & \multicolumn{2}{c}{-0.46} & dB \\
\rowcolor{lightgray}
Repetition rate & \multicolumn{2}{c}{11.8} & kHz \\
Electron $\pi$ pulse & \multicolumn{2}{c}{50} & ns \\
\rowcolor{lightgray}
Double excitation probability & 0.028 & 0.03&\\ 
Detector efficiency & \multicolumn{2}{c}{90} & \%\\
\bottomrule
\end{tabularx}
\caption{Parameters used for HOM and BK simulations. Each parameter was selected to mimic device properties; some from measurements and others based on fits. Parameters that are from fits are denoted by $^\star$.
}
\label{tab:Measured performance parameters}
\end{table}

\FloatBarrier

\appendix
\section{Contributors}\label{sec:authors}

\noindent
Francis Afzal$^1$, Mohsen Akhlaghi$^1$, Stefanie J. Beale$^1$, Olinka Bedroya$^1$, Kristin Bell$^1$, Laurent Bergeron$^1$, Kent Bonsma-Fisher$^1$, Polina Bychkova$^1$, Zachary M. E. Chaisson$^1$, Camille Chartrand$^1$, Chloe Clear$^{1}$, Adam Darcie$^1$, Adam DeAbreu$^1$, Colby DeLisle$^1$, Lesley A. Duncan$^1$, Chad Dundas Smith$^1$, John Dunn$^1$, Amir Ebrahimi$^1$, Nathan Evetts$^1$, Daker Fernandes Pinheiro$^1$, Patricio Fuentes$^1$, Tristen Georgiou$^1$, Biswarup Guha$^1$, Rafael Haenel$^1$, Daniel Higginbottom$^{1,2}$, Daniel M. Jackson$^1$, Navid Jahed$^1$, Amin Khorshidahmad$^1$, Prasoon K. Shandilya$^1$, Alexander T. K. Kurkjian$^1$, Nikolai Lauk$^1$, Nicholas R. Lee-Hone$^1$, Eric Lin$^1$, Rostyslav Litynskyy$^1$, Duncan Lock$^1$, Lisa Ma$^1$, Iain MacGilp$^1$, Evan R. MacQuarrie$^1$, Aaron Mar$^1$, Alireza Marefat Khah$^1$, Alex Matiash$^1$, Evan Meyer-Scott$^1$, Cathryn P. Michaels$^1$, Juliana Motira$^1$, Narwan Kabir Noori$^1$, Egor Ospadov$^1$, Ekta Patel$^1$, Alexander Patscheider$^1$, Danny Paulson$^1$, Ariel Petruk$^1$, Adarsh L. Ravindranath$^1$, Bogdan Reznychenko$^1$, Myles Ruether$^1$, Jeremy Ruscica$^1$, Kunal Saxena$^1$, Zachary Schaller$^1$, Alex Seidlitz$^1$, John Senger$^1$, Youn Seok Lee$^1$, Orbel Sevoyan$^1$, Stephanie Simmons$^{1,2}$, Oney Soykal$^1$, Leea Stott$^1$, Quyen Tran$^1$, Spyros Tserkis$^1$, Ata Ulhaq$^1$, Wyatt Vine$^1$, Russ Weeks$^1$, Gary Wolfowicz$^1$, Isao Yoneda$^1$\\

\noindent
$^1$ \textit{Photonic Inc.}\\
$^2$ \textit{Department of Physics, Simon Fraser University, Burnaby, British Columbia, Canada}

\end{document}